\documentclass{article}
\usepackage{amsmath,amssymb,amsfonts,amsthm}
\usepackage{color}
\usepackage{graphicx}
\usepackage{wrapfig}

\begin{document}

\flushbottom

\numberwithin{equation}{section}

\renewcommand{\figurename}{Fig.}
\def\refname{References}
\def\proofname{Proof}

\newtheorem{theorem}{Theorem}
\newtheorem{propos}{Proposition}
\newtheorem{remark}{Remark}

\def\tens#1{\ensuremath{\mathsf{#1}}}

\if@mathematic
   \def\vec#1{\ensuremath{\mathchoice
                     {\mbox{\boldmath$\displaystyle\mathbf{#1}$}}
                     {\mbox{\boldmath$\textstyle\mathbf{#1}$}}
                     {\mbox{\boldmath$\scriptstyle\mathbf{#1}$}}
                     {\mbox{\boldmath$\scriptscriptstyle\mathbf{#1}$}}}}
\else
   \def\vec#1{\ensuremath{\mathchoice
                     {\mbox{\boldmath$\displaystyle\mathbf{#1}$}}
                     {\mbox{\boldmath$\textstyle\mathbf{#1}$}}
                     {\mbox{\boldmath$\scriptstyle\mathbf{#1}$}}
                     {\mbox{\boldmath$\scriptscriptstyle\mathbf{#1}$}}}}
\fi

\newcommand{\diag}{\mathop{\rm diag}\nolimits}
\newcommand{\const}{\rm const}
\def\div{\operatorname{div}}

\begin{center}
{\Large\bf Fermi-like  acceleration and power-law energy growth in
nonholonomic systems}
			
\bigskip

{\large\bf I.\,A.~Bizyaev, A.\,V.~Borisov, V.\,V.~Kozlov, I.\,S.~Mamaev}

\end{center}

\begin{quote}
\begin{small}
\noindent
Steklov Mathematical Institute, Russian Academy of Sciences,
ul. Gubkina 8, Moscow, 119991 Russia

bizaev\_90@mail.ru, borisov@rcd.ru, kozlov@pran.ru and mamaev@rcd.ru

\end{small}

\bigskip

\begin{small}
\textbf{Abstract.}
This paper is concerned with a nonholonomic system with parametric excitation --- the Chaplygin sleigh
with time-varying mass distribution. A detailed analysis is made of the problem of the
existence of regimes with unbounded growth of energy (an analogue of Fermi's acceleration)
in the case where excitation is achieved by means of a rotor with variable angular momentum.
The existence of trajectories for which the translational velocity of the sleigh
increases indefinitely and has the asymptotics $\tau^{\frac{1}{3}}$ is proved. In addition,
it is shown that, when viscous friction with a nondegenerate Rayleigh function is added,
unbounded speed-up disappears and the trajectories of the reduced system
asymptotically tend to a limit cycle.

\smallskip



\end{small}
\end{quote}

\newtheorem{teo}{Theorem}
\newtheorem{rem}{Remark}
\newtheorem{lem}{Lemma}
\newtheorem{cor*}{Corollary}
\newtheorem{Def*}{Definition}
\newtheorem{pro*}{Proposition}
\newenvironment{Proof}

\newpage	
\section*{Introduction}
{\bf 1.} The Chaplygin sleigh on a plane is one of the best-known model systems in nonholonomic mechanics.
According to S.\,S.\,Chaplygin~\cite{Chaplygin}, the sleigh can be designed to have
a knife edge and two absolutely smooth legs attached to a rigid body.
In this case, the nonholonomic constraint is achieved by means of the knife edge: the translational velocity
at the point of contact of the knife edge is orthogonal to its plane (that is, to the
body-fixed direction).
A similar constraint can also be obtained by using a wheel pair \cite{HH} instead of a knife edge.

The free dynamics of the Chaplygin sleigh on a horizontal plane was studied by
C.\,Carath\'{e}odory \cite{Caratheodory}. Depending on the position of the center of mass
relative to the knife edge, the sleigh moves in a circle or asymptotically tends to
rectilinear motion. In the latter case, we have the classical scattering problem, for which the
scattering angle was found in \cite{Fuf}.
It is calculated explicitly, since the free motion of the sleigh is integrable and regular \cite{Caratheodory}.
The dynamics of the Chaplygin sleigh on an inclined plane is no longer integrable and exhibits
random asymptotic behavior depending on initial conditions \cite{BM2}.

The recent paper \cite{Jung} investigates the motion of the Chaplygin sleigh under the action of
random forces which model a fluctuating continuous medium. It turns out that in this case
the sleigh exhibits complex intricate behavior, which, according to the
authors, resembles random walks of bacterial cells with some diffusion component.
Similar behavior is exhibited by the sleigh under the action of periodic pulsed torque impacts,
which depend
on the orientation of the sleigh, and in the presence of viscous friction \cite{BorisovKuznetsov}.
In \cite{KozlovServoI,KozlovServoII}, the motion of the Chaplygin sleigh with servoconstraints
is explored. Other generalizations of the Chaplygin sleigh problem are discussed in \cite{bizyaevCylinder, JacobiIntegral}.

{\bf 2.} In this paper, we consider various aspects of the dynamics of a nonautonomous Chaplygin
sleigh
(i.e., with time-varying mass distribution). A detailed analysis is made
of the sleigh with a gyrostatic momentum periodically changing with time.
In practice this can be achieved by means of a rotor placed inside the body.

The possibility of self-propulsion of a rigid body in a fluid by means of an oscillating rotor
was predicted in \cite{KellyVortex}. The authors of \cite{KellyVortex} assert that the Kutta-Zhukovsky condition
is equivalent to a nonholonomic constraint,
 which is, generally speaking, incorrect from the viewpoint of physical principles of mechanics.
By the way, a nonholonomic model is also used in \cite{Fedorov,Vankerschaver1,Vankerschaver2} to describe the
motion of a plate in a fluid.
It should be noted that, when it comes to describing the motion of a rigid body in an ideal fluid, the equations
with nonintegrable constraints arise within the framework of vakonomic mechanics.
A detailed treatment of the problems concerning the scope of applicability of various
models with nonintegrable constraints is presented in \cite{BizyaevYMN}.

Our investigation of the dynamics of the Chaplygin sleigh with periodically time-dependent parameters
is closely related to the control problem. Since the sleigh can be designed to be a two-wheeled
robot \cite{HH}, it is of great practical importance, since the regimes arising at
fixed values of the angular velocities of eccentrics can be taken as basic regimes (called gaits),
which the body reaches after various maneuvers initiated by the control system. We note that
periodic changes in control functions were also considered in optimal control problems \cite{Leonard, Murray}.

{\bf 3.} In this paper, we examine in detail the dynamics of a reduced system
which decouples from
a complete system of equations and governs the evolution of the translational and angular
velocities of the sleigh.
From known solutions of a reduced system the dynamics of the point of contact is defined
by quadratures.

A reduced system is a system of two (nonlinear) first-order equations with
periodic coefficients which govern the evolution of the translational and angular velocities
of the sleigh.
However, in contrast to Hamiltonian systems with one and a half degrees of freedom, the
reduced system possesses no smooth invariant measure \cite{BM2}
 and can have different attractors (including strange ones) typical of dissipative systems.
In this sense, it is similar to various Duffing  and Van der Pol type oscillators with
parametric periodic excitation \cite{Sprott} and to the nonlinear Mathieu equation \cite{Izrailev, Ito}.
However, as noted in many publications, ``nonholonomic dissipation'', which
arises due to the divergence being sign-alternating, possesses specific features that require
additional research. Starting with \cite{b21-75}, strange attractors of different
nature \cite{BorisovTop,Gonchenko2013,Kuznetsov2012} are detected in nonholonomic systems.
A strange attractor for the Chaplygin sleigh with a material point, which executes periodic
oscillations in the direction transverse to the plane of the knife edge, is found in \cite{BizyaevKuznetsov,Bizyaev}.

{\bf 4.} The most interesting problem in the dynamics of the nonautonomous Chaplygin sleigh
is that of its speed-up (acceleration).
From a physical point of view, interest in it stems from the fact that unbounded growth of
energy and hence unbounded speed-up is achieved by means of a mechanism executing small, but regular oscillations.

As noted above, the system considered in this paper differs from Hamiltonian systems with one and a half
degrees of freedom. This difference is particularly pronounced in the situation with speed-up.

\begin{figure}[!ht]
	\begin{center}
		\includegraphics[totalheight=3.5cm]{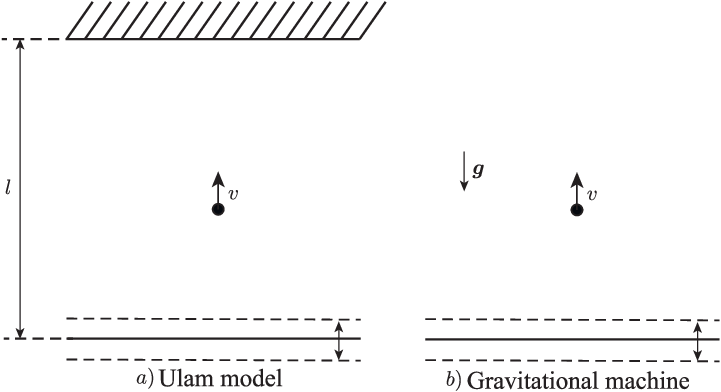}
	\end{center}
	\caption{}
	\label{fig60}
\end{figure}

The Hamiltonian speed-up model began to be discussed in the physical literature in connection with
the prediction of Fermi's acceleration \cite{Fermi} in the {\it Ulam model} \cite{Ulam}.
We recall that in this model the particle (in the absence of a gravitational field)
is located between two walls, with the lower wall moving periodically
in the vertical direction (see Fig. \ref{fig60}$a$).
This problem reduces to investigating an area-preserving two-dimensional Poincar\'{e} point map.
As shown numerically in \cite{Ulam} and then proved analytically in \cite{Zaslavskii, Brahic,LiebermanLi},
acceleration in different variations of the Ulam model is prevented by an invariant curve existing
at large velocities and predicted by KAM theory (see also \cite{Lichtenberg}).

If we remove the upper wall in the Ulam model and place the system in a gravitational field, then
we obtain the so-called {\it gravitational machine} (see Fig. \ref{fig60}$b$). In this problem
there exist trajectories for which the particle gathers speed without bound. For
a particular case of motion of the lower wall, the book \cite{Zaslav}
proposes the model of some random process for description of the dynamics.
Analysis of this process shows that the velocity increases as a function of time $t^{\frac{1}{3}}$. In the general
case, the presence of accelerating trajectories is proved by Pustylnikov in \cite{Pust}, where
it is also shown that the velocity of the particle at instants of collisions increases
as a function of time $t^{\frac{1}{2}}$.
Causes of the absence of an invariant curve at large velocities in a gravitational machine are discussed
in \cite{LichtenbergRevisited}.

Thus, in Hamiltonian systems with one and a half degrees of freedom the problem of
acceleration reduces to investigating the conditions under which
the KAM curves existing in the general case at large energies are destroyed.
Among modifications of the Ulam model in which acceleration is observed, we mention
generalizations associated with random \cite{Karlis,Hammersley} and piecewise
smooth \cite{Dolgopyat} motion of the wall and a relativistic generalization \cite{Pust_Rel}.

Acceleration in (nonlinear) natural Hamiltonian systems
 for two and a half and more degrees of freedom is closely related to Arnold's diffusion.
A detailed discussion of these issues can be found in \cite{RomKedar}.
We note that there are already a number of systems in which acceleration is shown
numerically \cite{Lenz, Pereira} or using analytical methods, which make it possible to prove
the presence of trajectories with increasing energy \cite{Bolotin, Koiller, Turaev}.
An insightful example, which has been intensively discussed recently, is
the two-dimensional periodically pulsating Birkhoff billiard. When pulsation is introduced,
different degrees of growth of energy of the particle are possible depending on the shape
of the boundary determining
the dynamical (stochastic, ergodic or regular) behavior of the ``frozen system''.
We note that the possibility of speeding up the particle by rotating the billiard is numerically
investigated in \cite{Costa}.
It is shown that, if there is a region on the boundary of the billiard in which the curvature
changes sign, acceleration is observed.

{\bf 5.} As noted above, nonholonomic systems possess no invariant
measure in the general case. Consequently, in the general case, KAM theory cannot be applied to them.
This is particularly clearly seen in the problem of the Chaplygin sleigh with periodically
changing gyrostatic momentum. It turns out that all solutions of the reduced system are
accelerating and have identical asymptotics of the growth of the translational velocity as a function of time $\tau^{\frac{1}{3}}$.

In the gravitational machine it is assumed that the particle collides absolutely elastically with
the wall. If one introduces dissipation into this system, assuming the impact to be not absolutely
elastic, then acceleration disappears in the gravitational machine \cite{Guk}. A similar
situation
is observed in the system considered in this paper.
If one introduces the force of viscous friction in the Chaplygin sleigh with variable
gyrostatic momentum, then unbounded speed-up disappears also.
In this paper we show that all trajectories of the reduced system tend to a limit cycle.

The problem dealt with in this paper shows that in nonholonomic mechanics acceleration is
characteristic even of small dimensions.
We mention the recent paper \cite{BizyaevKuznetsov}, in which the speed-up of the Chaplygin
sleigh is studied using the averaging method and the asymptotics of the degree of
speed-up as a function of time is obtained.
We note that the absence of an invariant measure turns out to be essential and necessary
for the presence of speed-up.
These issues are very important for developing the control of various mechanical devices.

The problem we consider here is a model problem, but its analysis allows one to pose the problem
of the possibility of speed-up in nonholonomic robots with a more complex control mechanism.
In particular, the control of spherical robots is discussed
in \cite{Pivovarova, ControlI, ControlII}.
The investigation of speed-up in such systems is a complicated and interesting problem.

In conclusion, we note that the Hamiltonian modification of the Chaplygin sleigh (the vakonomic sleigh)
 can be implemented by means of a plate moving in a fluid. If the mass distribution of the plate
depends on time, then the problem reduces to investigating the Hamiltonian system with one and a
half degrees of freedom.
In this case, it turns out impossible to speed up the plate indefinitely by
periodically changing the gyrostatic momentum.
This is prevented by an invariant КАМ curve existing at large velocities of motion~\cite{Ved}.

\newpage	

\section{Equations of motion}
The Chaplygin sleigh is a platform (rigid body) moving
on a horizontal plane with a nonholonomic constraint:
at some point the velocity  is always orthogonal to
a fixed direction.
This constraint can be obtained by means of a knife edge rigidly attached
to the platform \cite{Chaplygin}
or by means of a wheel pair \cite{HH}.
\begin{figure}[!ht]
	\begin{center}
		\includegraphics{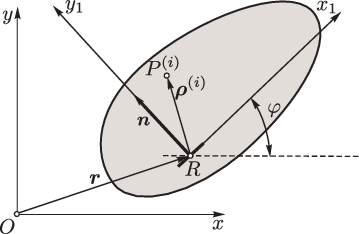}
	\end{center}
	\caption{The Chaplygin sleigh with moving points.}
	\label{fig01}
\end{figure}

To describe the motion of the sleigh, we define two
coordinate systems:
\begin{itemize}
	\item[---] a {\it fixed} (inertial) coordinate system
$Oxy$;
	\item[---] a {\it moving} coordinate system $Rx_1y_1$ attached to the platform.
\end{itemize}

We specify the position of the sleigh by the coordinates
$(x, y)$ of point $R$ in the fixed coordinate system $Oxy$,
and the orientation by the rotation angle $\varphi$.
Thus, the configuration space of the system
$\mathcal{N}=\{\boldsymbol{q} = (x, y, \varphi) \}$
coincides with the motion group of the plane $SE(2)$.

Let $\boldsymbol{v}=(v_1,v_2)$ denote the projections of the velocity
of point $R$ relative to the fixed coordinate system $Oxy$ onto the
moving axes $Rx_1y_1$
and let $\omega$ be the angular velocity of the body.
Then
\begin{equation}
\label{eq112}
\dot{x}=v_1\cos\varphi-v_2\sin\varphi, \quad \dot{y}=v_1\sin\varphi+v_2\cos\varphi, \quad \dot{\varphi}=\omega.
\end{equation}
The constraint equation in this case reads
\begin{equation}
\label{EquationSv}
v_2=0.
\end{equation}

Suppose that $n$ material points $P^{(i)}$, $i=1,..,n$ move on the
platform in a prescribed manner.
In this case the kinetic energy of the entire system can be represented
in the following form \cite{Bizyaev}:
\begin{eqnarray*}
T = \frac{1}{2}m\boldsymbol{v}^2 + m\omega(c_1(t)v_2  - c_2(t)v_1) + \frac{1}{2}(I(t) + m c^2_2(t))\omega^2  + \\+
m \big( v_1\dot{c}_1(t) + v_2\dot{c}_2(t) \big) + k(t)\omega, \\
\end{eqnarray*}
where $m$ is the mass of the entire system,  $I(t)$ is its
moment of inertia, $\boldsymbol{c}= (c_1(t), c_2(t))$ is the position
of the center of mass,
and $k(t)$ is the gyrostatic momentum arising from the
motion of the points.

For this system the Lagrange equations with undetermined
multipliers have the form
\begin{equation}
\begin{array}{ll}
\label{K05}
\displaystyle\frac{d}{dt}\left( \frac{\partial T}{\partial \omega} \right) = v_2 \frac{\partial T}{\partial v_1} - v_1 \frac{\partial T}{\partial v_2},\quad
\displaystyle\frac{d}{dt}\left( \frac{\partial T}{\partial v_1} \right) = \omega \frac{\partial T}{\partial v_2}, \\ [4 mm]
\displaystyle\frac{d}{dt}\left( \frac{\partial T}{\partial v_2} \right) = - \omega \frac{\partial T}{\partial v_1} + N,
\end{array}
\end{equation}
where $N$ is an undetermined multiplier which
is the reaction force at the point of contact $R$. This force
is directed transversely to the plane of the knife
edge.

In this case it is more convenient to represent this system
in the variables $(p, \omega)$, where $p$ is the
momentum given by the relation
$$
p = \left. \frac{\partial T}{\partial v_1}\right|_{v_2=0}     = m \Big(v_1 - c_2(t)\omega + \dot{c}_1(t) \Big).
$$

From the last equation of~(\ref{K05}) we find an expression
for the reaction force:
\begin{equation}
\begin{array}{ll}
\label{K03}
N=\left(1 - \frac{mc_1^2(t)}{I(t)} \right)\omega p + \left(\dot{c}_1(t) - \frac{c_1(t)}{I(t)}\big(\dot{I}(t) - mc_1(t)\dot{c}_1\big) \right)m\omega -\\  [3 mm]
\displaystyle - \frac{mc_1(t)}{I(t)}\left(m c_2(t)\ddot{c}_1(t) + \dot{k}(t) \right) + m \ddot{c}_2(t).
\end{array}
\end{equation}
Finally, from~(\ref{eq112}) and~(\ref{K05}) we obtain
equations of motion in the following form:
\begin{equation}
\begin{array}{ll}
\label{K01}
\dot{p}=mc_1(t)\omega^2 + m\omega \dot{c}_2(t), \\          [2 mm]
I(t)\dot{\omega}=-c_1(t)\omega p - \big(\dot{I}(t) - mc_1(t)\dot{c}_1(t) \big)\omega - mc_2(t)\ddot{c}_1(t) - \dot{k}(t), \\    [2 mm]
\dot{\varphi} = \omega, \quad  m\dot{x} = \big(p + c_2(t)\omega -\dot{c}_1(t)\big)\cos\varphi, \quad m\dot{y} = (p + c_2(t)\omega -\dot{c}_1(t))\sin\varphi.
\end{array}
\end{equation}

In~(\ref{K01}), the nonautonomous reduced system governing the evolution of
$(p, \omega)$ can be considered as a separate set.
Of great interest is the question of whether this system has trajectories unbounded on the plane $(p, \omega)$.
In this case the sleigh is observed to accelerate, that is,
the kinetic energy and hence  the velocity of the platform
must increase indefinitely with time.

As for acceleration of the sleigh, one should distinguish between
cases where the reaction force $N$ is a bounded and an unbounded
function of time.
Physically, unbounded increase in the reaction force $N$
implies that, at a certain instant of time,
slipping will start in the direction transverse to the plane
of the knife edge (i.e., the constraint~(\ref{EquationSv})
will be violated).

What is of interest from a practical point of view is
acceleration for which
the reaction force $N$ is a bounded function of time.

We consider separately several particular cases where the position of
the center of mass $\boldsymbol{c}(t)$, the moment of inertia $I(t)$ and
the gyrostatic momentum $k(t)$ depend on time.

{\bf Balanced case.} If the system is balanced relative to
the knife edge $c_1=0$ (i.e., the center of mass of the system
lies on the axis $Ry_1$), then the reduced system reduces to
a linear one.
In this case the equations of motion possess an additional
integral \cite{Bizyaev}:
\begin{equation}
\label{K02}
F=I(t)\omega + k(t).
\end{equation}
If we fix the level set of the integral $F=f$, then the equation
describing the momentum can be represented as
$$
\dot{p}=\frac{m\dot{c}_2(t)}{I(t)}\big( f - k(t)\big).
$$

In \cite{Bizyaev}, attention is given to the case in which
the functions $I(t)$, $k(t)$, $c_2(t)$ periodically depend on time.
Then, according to~(\ref{K02}), the angular velocity $\omega$
is also a periodic function of time, and momentum $p$
depends periodically on time or grows linearly with time.
In the latter case, acceleration is observed and, as follows
from~(\ref{K02}), the reaction force $N$ is an unbounded
function of time. Moreover, as numerical experiments
show, the trajectory of the point of contact in this case
has no directed motion~\cite{Bizyaev}.

\begin{figure}[!ht]
	\begin{center}
		\includegraphics{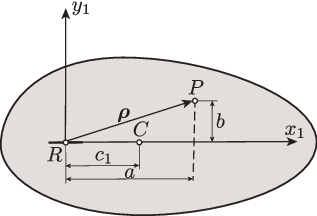}
	\end{center}
	\caption{The Chaplygin sleigh with
transversely oscillating mass.}
	\label{fig09}
\end{figure}

{\bf Transverse oscillations.} In~\cite{Bizyaev, BizyaevKuznetsov},
a detailed analysis is made of the case in which the center of mass
of the platform (point $C$) lies in the plane of the knife edge
and the material point executes oscillations transverse to
this plane (see Fig.~\ref{fig09}). Its position in the moving
coordinate system $Rx_1y_1$ is defined by the radius vector
$$
\boldsymbol{\rho}=\big(a,b\sin(\Omega t)\big).
$$
In this case we obtain
\begin{equation}
\begin{array}{ll}
I(t)=I_s + mb^2\mu(1-\mu)\sin^2(\Omega t), \quad k(t)=mba \mu\Omega\cos(\Omega t), \\ [2 mm]
c_1 ={\rm const }, \quad c_2(t)=\mu b\sin(\Omega t),
\end{array}
\end{equation}
where $I_s$ is the moment of inertia of the platform and
$\mu\in(0,1)$ is the ratio of the mass of the point to that
of the entire system.

In this case, the reduced system can be represented as
\begin{equation}
\begin{array}{ll}
\label{Pop}
\dot{p}\!=\!mc_1\omega^2 + m b\mu\Omega\cos(\Omega t) \omega, \\   [3 mm]
\displaystyle \dot{\omega}\!\!=\!\!-\frac{c_1\omega p +mb^2\mu(1-\mu)\Omega\sin(2\Omega t) \omega - mba \mu\Omega^2\sin(\Omega t)}{ I_s + mb^2\mu(1-\mu)\sin^2(\Omega t)}.
\end{array}
\end{equation}

Consider the quadratic function~\cite{Bizyaev}:
$$
F=ac_1p^2 + \big( (I_s + mb^2\mu(1-\mu)\sin^2(\Omega t))\omega +mba \mu\Omega\cos(\Omega t)  \big)^2.
$$
The derivative of this function along the system~(\ref{Pop})
has the form
$$
\dot{F}=-2c_1p\omega^2\big( mb^2\mu(1-\mu)\sin^2(\Omega t) + I_s - mac_1  \big).
$$
The conditions for which $\dot{F}$
in the half-planes $p>0$ and $p<0$ is sign-definite
are defined by the following relations:
\begin{equation}
\label{pp}
ac_1<0, \quad \mbox{or} \quad I_s - mac_1>0.
\end{equation}
In~\cite{Bizyaev} it is noted that, for parameters
satisfying~(\ref{pp}), one always observes acceleration during which
only momentum $p$ grows indefinitely with time (numerical
experiments show that it grows in proportion to
$\tau^{\frac{1}{3}}$). The first relation of~(\ref{pp}) has
a clear physical interpretation~--- the center of mass of the
system and the oscillating point
lie on different sides from the knife edge.

If relations~(\ref{pp}) are not satisfied, then one observes
chaotic oscillations and multistability for which all
trajectories of the reduced system~(\ref{Pop}) are bounded.
Also, on the Poincar\'{e} map one observes a strange attractor
which can coexist with invariant curves.

However, the conclusion on acceleration is not rigorously proved in this case.
The question of the behavior of the reaction force $N$ remains also open.
In addition, numerical experiments show that the point of contact has
directed motion.

\begin{figure}[!ht]
	\begin{center}
		\includegraphics{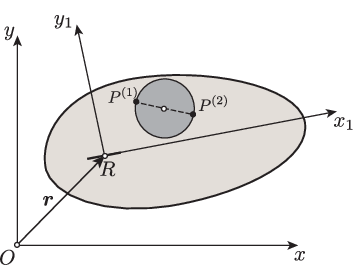}
	\end{center}
	\caption{The Chaplygin sleigh with gyrostatic momentum.}
	\label{Rot}
\end{figure}

{\bf A sleigh with a rotor.} In this paper we consider the motion of the
Chaplygin sleigh only with gyrostatic momentum, that is,
$I, c_1, c_2={\rm const}$. In this case, gyrostatic momentum
can be generated, for example, by two point masses rotating
about the common center (see Fig.~\ref{Rot}).

Assuming $c_1\neq 0$,
we define dimensionless variables and parameters:
\begin{equation}
\begin{array}{ll}
\label{eq_ob}
\alpha= \frac{c_1p}{I\Omega}, \quad u=\frac{\omega}{\Omega}, \quad \tau=\Omega t, \quad X=\frac{x}{c_1}, \quad Y=\frac{y}{c_1}, \\
\displaystyle A = \frac{mc_1^2}{I}\in(0,1), \quad  a=\frac{c_2}{c_1}, \quad \lambda=\frac{\dot{k}}{\Omega I},
\end{array}
\end{equation}
where $\Omega$ is some constant that has dimension
inverse to time.
The equations of motion in terms of these variables become
\begin{equation}
\label{eq011}
\frac{d \alpha}{d\tau}=Au^2, \quad \frac{d u}{d\tau}=-\alpha u  - \lambda(\tau),
\end{equation}
\begin{equation}
\label{eq011_2}
\frac{d \varphi}{d\tau}=u, \quad  \frac{d X}{d\tau}=\left( \frac{\alpha}{A} + a u \right)  \cos\varphi, \quad
\frac{d Y}{d\tau}= \left( \frac{\alpha}{A} + a u \right)   \sin\varphi.
\end{equation}

We represent the relation for the reaction force in
dimensionless variables in the form
$$
\Lambda=\frac{N}{ma\Omega^2}=\left( \frac{1}{A} - 1\right)  \alpha u - \lambda(\tau).
$$

\begin{rem}
Let the gyrostatic momentum $k(t)$ and the moment of inertia
$I(t)$ depend on time and let the center of mass be fixed,
$\boldsymbol{c}={\rm const}$. We show that in this case the reduced
system can be represented in the form~(\ref{eq011}), where
$A$ is a positive function of time.

Indeed,  the reduced system has the form
$$
\dot{p}=\frac{mc_1}{I(t)^2}M^2, \quad \dot{M}=-\frac{c_1}{I(t)}M p - \dot{k},
$$
where $M=I(t)\omega$.

Using the fact that the moment of inertia $I(t)$ is a positive
function, we rescale time as
$$
\tau=\Omega\int \frac{mc_1^2}{I(t)}dt
$$
 and introduce new variables
$$
\alpha=\frac{p}{mc_1\Omega}, \quad u=\frac{M}{mc_1^2\Omega}.
$$
The equations of motion in this case can be represented as
\begin{eqnarray*}
\frac{d \alpha}{d\tau}=A(\tau)u^2, \quad \frac{d u}{d\tau}=-\alpha u  - \lambda(\tau), \\
\lambda(\tau)=\frac{d k}{d\tau}, \quad
A(\tau) = \frac{mc_1^2}{I(\tau)}.
\end{eqnarray*}

\end{rem}

\newpage	
\section{Proof of the existence of nonlinear acceleration}
Consider in more detail the question of ``constant''
acceleration of the sleigh by means of a rotor.
This question reduces to investigating the possibility of
existence of unbounded trajectories for the reduced
system~(\ref{eq011}).
We note that parameter $A$ in this system can be eliminated
by the transformation
\begin{equation}
\label{zz}
 \quad u \to \frac{1}{\sqrt{A}}u,\quad \lambda(\tau)\to \frac{1}{\sqrt{A}}\lambda(\tau),
\end{equation}
after which the equations of motion can be represented as
\begin{equation}
\label{eq5}
\alpha'=u^2, \quad u'=-\alpha u - \lambda(\tau),
\end{equation}
where the prime denotes the derivative with respect to $\tau$.

The right-hand sides of the system of differential equations~(\ref{eq5}) contain quadratic terms.
Solutions of such systems can go to infinity in finite time.
The simplest example of a system with this property is $x'=x^2$,
$x\in \mathbb{R}$.

We show that all solutions of the system~(\ref{eq5}) are
defined on the whole axis of new time $\mathbb{R}=\{\tau\}$.
For this purpose we consider the function
$$
V=\frac{1}{2}(u^2 + \alpha^2).
$$
By virtue of~(\ref{eq5}) the derivative of this function is
$$
V'=-u\lambda(\tau).
$$
Hence,
$$
|V'|=|u||\lambda|\leqslant \sqrt{2}\sqrt{V}|\lambda|.
$$
Since $V\geqslant 0$ and $\sqrt{V}\geqslant 0$, it follows that
$$
|V^{-\frac{1}{2}}V'| \leqslant \sqrt{2}|\lambda|,
$$
or
$$
|2\big(V^{\frac{1}{2}} \big)'| \leqslant  \sqrt{2}|\lambda|.
$$
Integrating the inequalities
$$
-\frac{1}{\sqrt{2}}|\lambda(\tau)|\leqslant - |\big(V^{\frac{1}{2}} \big)' |\leqslant \big(V^{\frac{1}{2}} \big)'\leqslant
|\big(V^{\frac{1}{2}} \big)' |\leqslant\frac{1}{\sqrt{2}}|\lambda(\tau)|
$$
in the interval from $0$ to $\tau$, we find that
$$
|V^{\frac{1}{2}}(\tau) - V^{\frac{1}{2}}(0) |\leqslant\frac{1}{\sqrt{2}}\int_0^{\tau}|\lambda(s)|ds.
$$
Consequently, the function $V(\tau)$ (along with the function $\alpha(\tau)$
and $u(\tau)$) can tend to $+\infty$
only as $|\tau|\to \infty$, which is the required result.

\begin{teo}
	\label{teo1}
	Assume that the functions $\lambda(\tau)$ and $\lambda'(\tau)$
are bounded and the function $\lambda(\tau)$ does not tend to
zero as $\tau\to
	+\infty$. If at the initial instant of time $\alpha>0$,
then for any initial value of the variable $u$
	\begin{itemize}
		\item[{1)}] $\alpha(\tau)$ tends to $+\infty$ as
$\tau\to+\infty$,
		\item[{2)}] $u(\tau)\to 0$ as $\tau\to +\infty$,
		\item[{3)}] the function $\alpha(\tau)u(\tau)$
is bounded,
		\item[{4)}] $u'(\tau)\to 0$ as $\tau\to +\infty$.
	\end{itemize}
\end{teo}

In particular, under these conditions
$\alpha'(\tau)\to 0$ as $\tau\to +\infty$ and the constraint reaction
$\Lambda$ is bounded. The conditions of Theorem~\ref{teo1}
for the function $\lambda$ hold if $k(\tau)$
is a periodic or (more generally)
conditionally periodic nonconstant function of time.
\proof[Proof of Theorem~\ref{teo1}]
Conclusion 1 is proved using the following lemma.
\begin{lem}[Hadamard~\cite{Hadamard}]
	If the function $f(\tau)$ tends to the limit as
$\tau\to +\infty$ and the functions $f'(\tau)$ and $f''(\tau)$
are bounded, then $f'(\tau)\to
	0$ as $\tau\to +\infty$.
\end{lem}
The monotone growth of the function $\alpha(\tau)$ follows from
the fact that the derivative $\alpha'$ is positive for
$u\ne 0$ and that $u=0$ is not an invariant submanifold of the
system~(\ref{eq5}).

Assume that the function $\alpha(\tau)$ is bounded.
Then
$$
\lim\limits_{\tau\to\infty} \alpha(\tau)=\bar{\alpha}>0.
$$
The positiveness of the limit follows from the assumption that
$\alpha(0)>0$.

We show that in this case the function $u(\tau)$ is bounded.
Indeed, it satisfies the linear differential equation
\begin{equation}
\label{eq6}
\frac{du}{d\tau}=-\alpha(\tau) u-\lambda(\tau).
\end{equation}
We solve it by the method of variation of constants.
The solution of the homogeneous equation is
$$
u=C e^{-\int\limits_0^\tau\alpha(s)ds}.
$$
Now, assuming $C$ to be a function of $\tau$, we obtain
$$
C'=-\lambda(\tau) e^{\int\limits_0^\tau \alpha(s)ds}.
$$
Let us introduce a new function $\widetilde{C}$
by the following formula:
$$
\widetilde{C}'=-\mu e^{\int\limits_0^\tau \alpha(s)ds}, \quad \mu=\sup |\lambda(\tau)|.
$$
It is clear that $|\widetilde{C}'|\geqslant |C'|$.
Consequently, if $C(0)=\widetilde{C}(0)$, then
\begin{equation}
\label{eq70}
|C(\tau)|\leqslant |\widetilde{C}(\tau)|
\end{equation}
for all $\tau\geqslant 0$.

Let us calculate
$$
\lim\limits_{\tau\to +\infty} \widetilde{C}(\tau) e^{-\int\limits_0^\tau \alpha(s)ds}=\lim\limits_{\tau\to+ \infty}\frac{\widetilde{C}(\tau)}{e^{\int\limits_0^\tau\alpha(s)ds}}.
$$
Since $\alpha(s)\to \bar{\alpha}>0$, the denominator of this fraction
tends to $+\infty$ (as does the absolute value of the numerator).
Therefore, one can use L'H\^{o}pital's rule: this limit is
$$
\lim\limits_{\tau\to+\infty}\frac{\widetilde{C}'}{\alpha(\tau) e^{\int\limits_0^\tau\alpha(s)ds}}=-\frac{\mu}{\bar{\alpha}}.
$$
Thus (according to~(\ref{eq70})), the function
$|u(\tau)|=\Big|C(\tau) e^{-\int\limits_0^\tau\alpha(s)ds}\Big|\leqslant \Big|\widetilde{C}(\tau)
e^{-\int\limits_0^\tau\alpha(s)ds}\Big|$ is bounded.

Further, according to~(\ref{eq5}),
$$
\alpha''=2uu'=-2\alpha u^2-2u\lambda(\tau)
$$
is also bounded. Therefore (by Hadamard's lemma),
$\alpha'(\tau)\to 0$. But then (according to~(\ref{eq5}))
$u(\tau)\to 0$ as $\tau\to
+\infty$.

We now consider the second derivative
$$
u''=-\alpha'u-\alpha u'+\lambda'(\tau).
$$
This derivative is bounded since (according to~(\ref{eq5}))
$u'$ is bounded. Since \ $\!u(\tau)\to 0$ and the derivatives
$u'(\tau)$ and $u''(\tau)$ are bounded,(again according to
Hadamard's lemma) $u'(\tau)\to 0$ as $\tau\to+\infty$.
But this contradicts the second equation of~(\ref{eq5}), since
$\alpha(\tau)\to \bar{\alpha}$ (by assumption), $u(\tau)\to 0$, and
the function $\lambda(\tau)$ does not tend to zero as $\tau\to +\infty$.

The resulting contradiction proves conclusion 1. Next, the following lemma is needed.
\begin{lem}
	\label{lem2}
	Under the conditions of Theorem~\ref{teo1} the
relation
	\begin{equation}
	\label{eq2}
	\frac{\int\limits_0^\tau f(p)dp}{f(\tau)}, \quad {\rm where} \quad f(p)=e^{\int\limits_0^p \alpha(s)ds},
	\end{equation}
	tends to zero as $\tau\to +\infty$.
\end{lem}
\proof Since the denominator of~(\ref{eq2}) tends to infinity
as $\tau\to +\infty$, one can use L'H\^{o}pital's rule
and the statement (already proved) that $\alpha(\tau)\to +\infty$ as
$\tau\to+\infty$.
\qed

We now prove conclusion 2. The function $u(\tau)$ as a
solution of the linear inhomogeneous differential equation
$$
u'=-\alpha(\tau)u-\lambda(\tau)
$$
is the sum of two functions
\begin{equation}
\label{eq3}
u(0)e^{-\int\limits_0^\tau \alpha (s)ds}
\end{equation}
and
\begin{equation}
\label{eq4}
-e^{-\int\limits_0^\tau \alpha(s)ds}\int\limits_0^\tau \lambda(p)e^{\int\limits_0^p \alpha(s)ds}dp.
\end{equation}
The function~(\ref{eq3}) tends superexponentially fast to zero
as $\tau\to +\infty$. Since $\lambda(\tau)$
is bounded, the function~(\ref{eq4}) also tends to zero according to
Lemma~\ref{lem2}.

To prove conclusion 3, we make use of formulae~(\ref{eq3})
and~(\ref{eq4}), the sum of which is the function
$u(\tau)$. The product of $\alpha(\tau)$ and the
function~(\ref{eq3}) tends to zero as $\tau\to +\infty$. Indeed,
according to L'H\^{o}pital's rule, the limit of this product is equal to $$
u(0)\lim\limits_{\tau\to +\infty}\frac{\alpha'(\tau)}{\alpha(\tau)e^{\int\limits_0^\tau\alpha(s)ds}}=0,
$$
since $\alpha'(\tau)\to 0$ (according to conclusion 2) and $\alpha(\tau)\to \infty$ as $\tau\to +\infty$.

Since the function $\lambda(\tau)$ is bounded, the product of
$\alpha(\tau)$ and the function~(\ref{eq4}) is estimated from above
by the function
$$
\frac{\alpha(\tau)\int\limits_0^\tau f(p)dp}{f(\tau)}, \quad {\rm where} \quad f(p)=e^{\int\limits_0^p\alpha(s)ds}.
$$
According to L'H\^{o}pital's rule, the limit of this function as $\tau\to+\infty$ is equal to
$$
\lim\limits_{\tau\to +\infty}\frac{\alpha'\int\limits_0^\tau f(p)dp+\alpha(\tau)f(\tau)}{\alpha(\tau)f(\tau)}=1
$$
(by Lemma~\ref{lem2}). Consequently, the product
$\alpha(\tau)u(\tau)$ is indeed bounded.

According to the second equation of~(\ref{eq5}), the derivative
$u'(\tau)$ is bounded. We prove that $u'(\tau)\to 0$ as
$\tau\to
+\infty$.

Indeed, according to~(\ref{eq3}) and~(\ref{eq4}),
\begin{equation}
\label{eq5_0}
-u'=\alpha u+\lambda=\frac{\alpha\Big[u(0)-\int\limits_0^\tau\lambda(p)f(p)dp\Big]+\lambda(\tau)f(\tau)}{f(\tau)},
\end{equation}
where $f$ is given by formula~(\ref{eq2}). It is clear that
$$
u(0)\frac{\alpha(\tau)}{f(\tau)}\to 0
$$
as $\tau\to +\infty$. We calculate the limit of the two remaining terms
in~(\ref{eq5_0}) again by L'H\^{o}pital's rule. It is equal to
\begin{equation}
\label{eq6_0}
\lim\limits_{\tau\to +\infty}\frac{-\alpha'\int\limits_0^\tau \lambda f dp+\lambda' f}{\alpha f}.
\end{equation}
Taking into account the assumption about
boundedness of the functions $\lambda(\tau)$ and
$\lambda'(\tau)$,
as well as the already established properties $\alpha(\tau)\to
+\infty$, $\alpha'(\tau)\to 0$ and Lemma~\ref{lem2}, we find that
the limit~(\ref{eq6_0}) is zero. This proves conclusion 4.

\qed

\begin{teo}
	\label{teo2}
	Suppose that the conditions of Theorem~\ref{teo1} hold and
there exists the average
	\begin{equation}
	\label{eq7}
	\langle\lambda^2\rangle=\lim\limits_{\tau\to+ \infty}\frac{1}{\tau}\int\limits_0^\tau\lambda^2(p)dp.
	\end{equation}
	Then, as $\tau\to+ \infty$,
	\begin{equation}
	\label{eq8}
	\frac{\alpha^3(\tau)}{3}=\langle\lambda^2\rangle\tau+o(\tau).
	\end{equation}
\end{teo}

The average value of~(\ref{eq7}) exists if $\lambda$
is a periodic or conditionally periodic function of time.
	If $\lambda$ does not vanish, then, obviously,
$\langle\lambda^2\rangle$ is positive. Formula~(\ref{eq8})
can be represented in the following equivalent form:
	$$
	\alpha=\sqrt[3]{3\langle\lambda^2\rangle}\tau^{\frac{1}{3}}+o\Big(\tau^{\frac{1}{3}}\Big).
	$$
\proof [Proof of Theorem~\ref{teo2}]	
 Since $u'(\tau)\to 0$ as $\tau\to +\infty$ (conclusion 4 of Theorem~\ref{teo1}),
according to the second equation of~(\ref{eq5}) we have
$$
\alpha u=-\lambda(\tau)+f(\tau),
$$
where $f(\tau)= o(1)$. Hence,
$$
u=\frac{-\lambda+f}{\alpha}.
$$
Since $\alpha'=u^2$, it follows that
\begin{equation}
\label{eq9}
\Big[\frac{1}{3}\alpha^3\Big]'=\lambda^2-2\lambda f+f^2.
\end{equation}
Let $g(\tau)$~be one of the functions
$$
-2\lambda(\tau)f(\tau) \quad {\rm or} \quad f^2(\tau).
$$
It is clear that $g(\tau)\to 0$ as $\tau\to +\infty$.
Consequently,
$$
\int\limits_0^\tau g(s)ds=o(s).
$$
Indeed, since $g(\tau)=o(1)$, we have
$$
\lim\limits_{\tau\to +\infty}\frac{1}{\tau}\int\limits_0^\tau g(s)ds=0.
$$

Integrating~(\ref{eq9}), we obtain the required formula:
$$
\frac{\alpha^3}{3}=\langle\lambda^2\rangle\tau+o(\tau).
$$
\begin{teo}
	\label{teo3}
	Suppose that the conditions of Theorem~\ref{teo1} hold,
$u(0)=0$ and the analytic function $k(\tau)$ is such that
for any $a>0$ one can find the root of the equation
	$$
	k(\tau)=k(a),
	$$
	which is strictly larger than $a$. Then the function
$u(\tau)$ has infinitely many zeros as $\tau\to +\infty$.
	
\end{teo}

{\it Proof of Theorem~\ref{teo3} uses the formula}

\begin{equation}
\label{eq11}
u(\tau)=-\frac{1}{f(\tau)}\int\limits_0^\tau\lambda(p)f(p)dp,
\end{equation}
and the property of strict monotone growth of the function $f$.
The infinity of the number of zeros of the function~(\ref{eq11}) follows from
the following general result~\cite[Sec.~6]{KozlovTor}.

\begin{lem}
	\label{lem3}
	Let $\tau_1$~be the first positive zero of the analytic
function
	\begin{equation}
	\label{eq12}
	\int\limits_a^\tau\lambda(t)dt
	\end{equation}
	and let $f$~be a positive nondecreasing function. Then
the integral
	$$
	\int\limits_a^\tau\lambda(t)f(t)dt
	$$
	vanishes on the interval $(a, \tau_1]$.
\end{lem}

In our case, in view of~(\ref{eq_ob}) and~(\ref{zz}),
the integral~(\ref{eq12}) is
$$
B(k(\tau)-k(a)), \quad B=\frac{1}{\sqrt{A}I\Omega}
,$$ and therefore the conclusion of Theorem~\ref{teo3} follows from
formula~(\ref{eq11}) and Lemma~\ref{lem3}.

\begin{cor*}
	Suppose that the conditions of Theorem~\ref{teo1} hold,
$u(0)=0$, and the analytic function $k(\tau)$ is
periodic with period $T$. Then
	the function $u(\tau)$ has infinitely many zeros as
$\tau\to  +\infty$, and the distance between its neighboring
zeros does not exceed $T$.
\end{cor*}

\newpage
\section{Dynamics in configuration space}
\label{DK}
As shown in the previous section
for solutions to the reduced system, in view of the transformation~(\ref{zz}) we have
\begin{equation}
\begin{array}{ll}
\label{Ad_02}
\alpha(\tau) = \sigma \tau^{\frac{1}{3}} + o(\tau^{\frac{1}{3}}), \qquad
u(\tau) = - \frac{\lambda(\tau)}{\sigma}\tau^{-\frac{1}{3}} + o(\tau^{-\frac{1}{3}}), \\
\sigma = \left( 3 A\langle\lambda^2\rangle\right)^{\frac{1}{3}},
\end{array}
\end{equation}
where $\langle\lambda^2\rangle$ is defined by~(\ref{eq7}).

From the given solutions of the reduced system the orientation of the
Chaplygin sleigh and the motion of the point of contact
are defined by quadratures~(\ref{eq011_2}).
If we restrict ourselves to the asymptotics~(\ref{Ad_02}), we obtain
\begin{equation}
\begin{array}{ll}
\label{Ad_01}
\displaystyle \frac{d \varphi}{d\tau}=- \frac{\lambda(\tau)}{\sigma}\tau^{-\frac{1}{3}}, \\          [4 mm]
\displaystyle \frac{d X}{d\tau}=\frac{\sigma}{A} \tau^{\frac{1}{3}} \cos\varphi, \quad
\displaystyle \frac{d Y}{d\tau}= \frac{\sigma}{A} \tau^{\frac{1}{3}} \sin\varphi.
\end{array}
\end{equation}

We let $\tau=\tau_0>0$ denote time from which the asymptotics~(\ref{Ad_02})
describes ``well'' the solution of the reduced system, and supplement the system~(\ref{Ad_01})
with the following initial conditions:
\begin{equation}
\label{Ad_001}
\varphi(\tau_0)=X_0, \quad X(\tau_0)=Y_0, \quad Y(\tau_0)=Z_0.
\end{equation}

Consider the system~(\ref{Ad_01}) in more detail in two cases:
\begin{itemize}
	\item[1)]  {\it constantly accelerating rotor}~--- the angular velocity of the rotor is
a linear function of time, in this case $\lambda(\tau)=\lambda_0={\rm const}$;
	\item[2)]  {\it periodically oscillating rotor}~--- the angular velocity of the rotor is
a periodic function of time.
	As an example we consider the function $\lambda(\tau)=\lambda_0\sin\tau$.
\end{itemize}	


{\bf  Constantly accelerating rotor ($\lambda(\tau)=\lambda_0>0$).}
Explicitly integrating the system~(\ref{Ad_01}), we find
\begin{equation}
\varphi(\tau)=-s_1(\tau^{\frac{2}{3}} - \tau_0^{\frac{2}{3}}) + \varphi_0, \quad s_1=\left( \frac{9\lambda_0 }{8A}\right)^{\frac{1}{3}},
\end{equation}
\begin{equation}
\begin{array}{ll}
\label{conp01}
X(\tau) = 2\big(\cos\varphi(\tau) - s_1 \tau^{\frac{2}{3}}\sin\varphi(\tau)\big) -
2\big(\cos\varphi_0 - s_1 \tau_0^{\frac{2}{3}}\sin\varphi_0  \big) + X_0, \\  [2 mm]
Y(\tau) = 2\big( \sin\varphi(\tau) + s_1\tau^{\frac{2}{3}}\cos\varphi(\tau)\big) -
2\big(\sin\varphi_0 + s_1 \tau_0^{\frac{2}{3}}\cos\varphi_0   \big) + Y_0.	
\end{array}
\end{equation}

\begin{figure}[!ht]
	\begin{center}
		\includegraphics[totalheight=4.5cm]{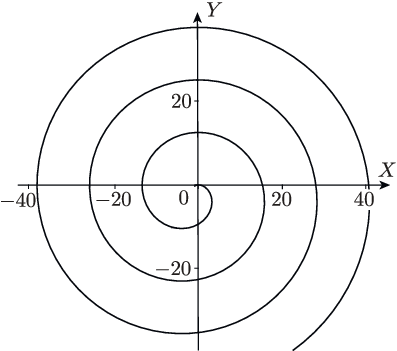}
	\end{center}
	\caption{The curve~(\ref{conp01}) with fixed $A=1$, $\lambda_0=1$, $\varphi_0=0$, $X_0=0$, $Y_0=0$, $\tau_0=1$.}
	\label{fig07}
\end{figure}

As we see, the trajectory of the point of contact is an untwisting spiral (Fig. \ref{fig07})
and in this case there is no directed motion of the sleigh.

\begin{figure}[!ht]
	\begin{center}
		\includegraphics[totalheight=9.5cm]{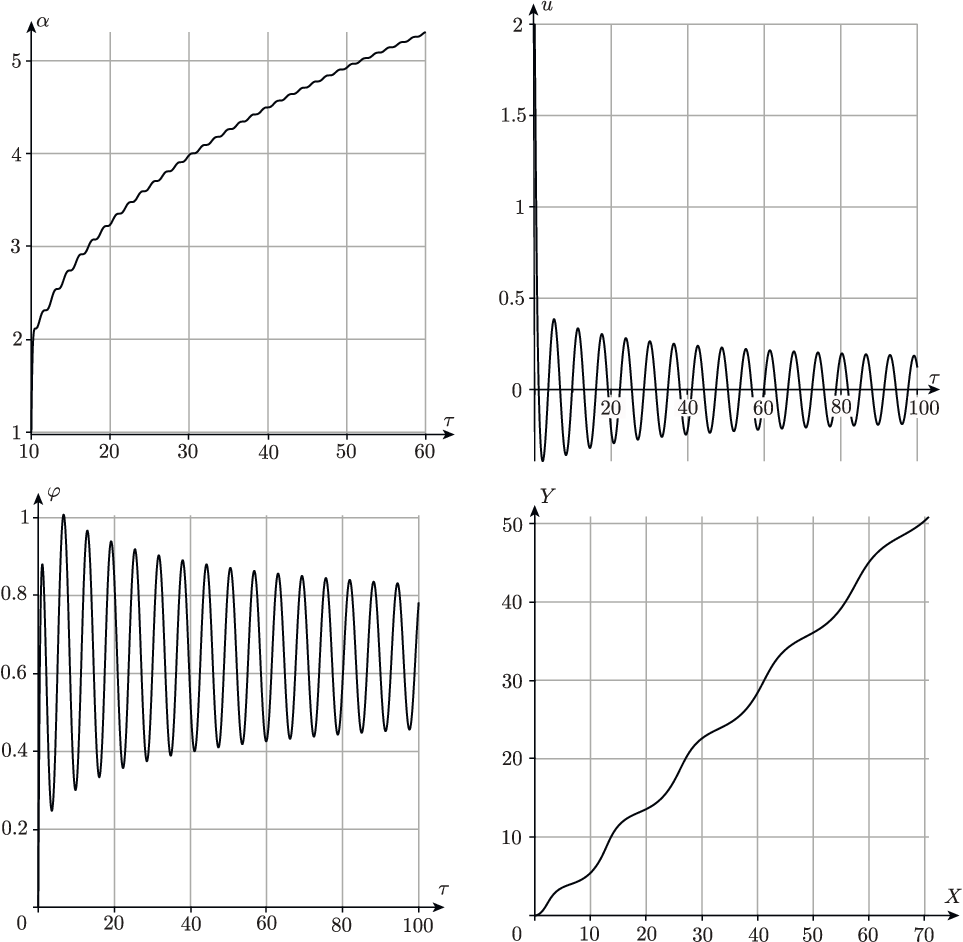}
	\end{center}
	\caption{Functions $\alpha(\tau)$, $u(\tau)$, $\varphi(\tau)$, $Y(X)$ plotted
numerically for fixed parameters $\lambda_0=1$, $A=1$ and initial conditions
$\tau=0$, $\alpha=1$,  $u=2$,
		$\varphi=0$, $X=0$, $Y=0$.}
	\label{fig15}
\end{figure}

{\bf Periodically oscillating rotor ($\lambda(\tau)=\lambda_0\sin\tau$).}
A typical behavior of the solutions to the system~(\ref{eq011}) and~(\ref{eq011_2}) in this case
is presented in Fig.~\ref{fig15}. As numerical experiments show, the following statement holds.

{\it In the case of a periodically oscillating rotor the angle of rotation of the sleigh
$\varphi$ tends to a finite limit, and the ``limit'' motions of the point of contact are
oscillations (with constant amplitude) in a neighborhood of a straight line. }

In order to show the validity of this statement, we consider
the system~(\ref{Ad_01}) in the approximation~(\ref{Ad_02}) and
make use of the relation~\cite[p.401]{Oldham}
$$
\int_\tau^\infty \frac{\sin\xi}{\xi^{\frac{1}{3}}}d\xi = \frac{\cos\tau}{\tau^{\frac{1}{2}}}+O(\tau^{-\frac{4}{3}}).
$$
This yields
\begin{eqnarray*}
\varphi(\tau) = \widetilde{\varphi}+ s_2 \frac{\cos\tau}{\tau^{\frac{1}{3}}} + O(\tau^{-\frac{4}{3}}), \quad
s_2=\left( \frac{2\lambda_0}{3A} \right)^{\frac{1}{3}}, \\
\widetilde{\varphi}=\varphi_0 - s_2\frac{\cos\tau_0}{\tau_0^{\frac{1}{3}}} + O(\tau_0^{-\frac{4}{3}}),
\end{eqnarray*}
that is, the angle of rotation of the sleigh tends to a fixed value of~$\widetilde{\varphi}$
as $t\to+\infty$.
We substitute the function $\varphi(\tau)$ obtained into the equations for the evolution of
the point
of contact
and expand the resulting expressions in powers of $\tau^{-\frac{1}{3}}$.
Neglecting terms of order $O(\tau^{-\frac{4}{3}})$ and $O(\tau_0^{-\frac{4}{3}})$,
we represent the equations of motion for the point of contact in the form
\begin{equation*}
\begin{array}{ll}
\displaystyle \frac{dX}{d\tau} = \frac{\sigma}{A}\left( \tau^{\frac{1}{3}}\cos\widetilde{\varphi} + s_2 \cos\tau \sin \widetilde{\varphi}\right), \\    [3 mm]
\displaystyle \frac{dY}{d\tau} = \frac{\sigma}{A}\left( \tau^{\frac{1}{3}}\sin\widetilde{\varphi} - s_2 \cos\tau \cos \widetilde{\varphi}\right).
\end{array}
\end{equation*}
	
If we rotate the fixed coordinate system through angle $\widetilde{\varphi}$:
$$
\widetilde{X}=X\cos\widetilde{\varphi} + Y\sin \widetilde{\varphi}, \quad
\widetilde{Y}=-X\sin\widetilde{\varphi} +Y\cos\widetilde{\varphi},
$$
then the equations of motion become
$$
\frac{d\widetilde{X}}{d\tau}=\frac{\sigma}{A}\tau^{\frac{1}{3}}, \quad
\frac{d\widetilde{Y}}{d\tau}=-s_2\frac{\sigma}{A} \cos\tau.
$$
Explicitly integrating this system, we find
\begin{equation}
\begin{array}{ll}
\label{App}
\displaystyle \widetilde{X}(\tau) = \widetilde{X}_0 + \frac{3}{4}\frac{\sigma}{A}(\tau^{\frac{4}{3}} - \tau_0^{\frac{4}{3}}),  \\   [3 mm]
\displaystyle \widetilde{Y}(\tau) = \widetilde{Y}_0 - \frac{\sigma}{A} s_2(\sin\tau - \sin\tau_0),
\end{array}
\end{equation}
where $\widetilde{X}_0$ and $\widetilde{Y}_0$ are initial conditions calculated
using~(\ref{Ad_001}).
%
Thus, along the axis $\widetilde{X}$ the sleigh moves away in proportion to $\tau^{\frac{4}{3}}$,
and along the axis $\widetilde{Y}$ it executes oscillations with constant amplitude
$\displaystyle\frac{\sigma s_2}{A}=\frac{\lambda_0}{A}$.

Let us compare relation~(\ref{App}) with the numerical solution of the system~(\ref{eq011})--(\ref{eq011_2}).
To do so, we calculate the angle:
\begin{equation}
\begin{array}{ll}
\label{eq_phi}
\displaystyle \tilde{\varphi}_n=\frac{\psi_1 + \psi_2}{2}, \quad \psi_1=\frac{1}{2\pi} \int\limits_{\tau_0}^{\tau_0 + 2\pi n}\varphi(\tau)d\tau, \\ [3 mm]
\displaystyle \psi_2=\frac{1}{2\pi} \int\limits_{\tau_0 + \pi}^{\tau_0 + \pi(2n+1)}\varphi(\tau)d\tau, \quad n\in \mathbb{N},
\end{array}
\end{equation}
where the function $\varphi(\tau)$ is a numerical solution of the initial system.
The angle of rotation is defined by
$$
\tilde{\varphi}=\lim\limits_{n\to\infty}\varphi_n,
$$
that is, the larger $n$, the more exactly $\varphi_n$ approximates $\tilde{\varphi}$.

\begin{figure}[!ht]
	\begin{center}
		\includegraphics[totalheight=4cm]{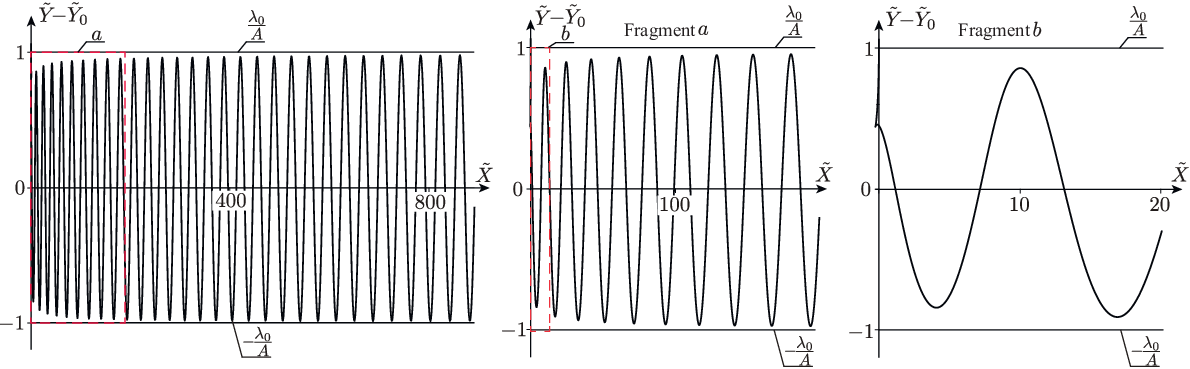}
	\end{center}
	\caption{Trajectory of the point of contact plotted numerically for fixed parameters
$\lambda_0=1$, $A=1$ and initial conditions $\tau=0$, $\alpha=0$, $u=0$, $\varphi=0$, $X=0$, $Y=0$. In this case   $\tilde{\varphi}=4.638$ ($n=10$) and $\tilde{Y}_0=-1.068$.}
	\label{fig20}
\end{figure}

A typical view of the trajectory of the point of contact is shown in Fig.~\ref{fig20}.
As we see, the relation obtained for the amplitude agrees well with the numerical experiments.
However, a rigorous proof of this fact requires investigating more detailed estimates for the
functions $\alpha(\tau)$ and
$u(\tau)$ (than that considered in this section~(\ref{Ad_02})).

%

\newpage
\section{Viscous friction}

We assume that the motion of the sleigh occurs in the presence of viscous friction force with
a dissipative Rayleigh function of the form
$$
F=\frac{1}{2}\big(\nu_1 v_1^2+\nu_2 \omega^2\big).
$$

For the system of Lagrange equations with
undetermined multipliers we write
\begin{equation*}
\begin{array}{ll}
\displaystyle\frac{d}{dt}\left( \frac{\partial T}{\partial v_1} \right) = \omega \frac{\partial T}{\partial v_2} - \nu v_1, &
\displaystyle\frac{d}{dt}\left( \frac{\partial T}{\partial \omega} \right) = v_2 \frac{\partial T}{\partial v_1} - v_1 \frac{\partial T}{\partial v_2}-\nu_2 \omega, \\[4mm]
\displaystyle\frac{d}{dt}\left( \frac{\partial T}{\partial v_2} \right) = - \omega \frac{\partial T}{\partial v_1} + N.
\end{array}
\end{equation*}

\begin{figure}[!ht]
	\begin{center}
		\includegraphics[totalheight=5cm]{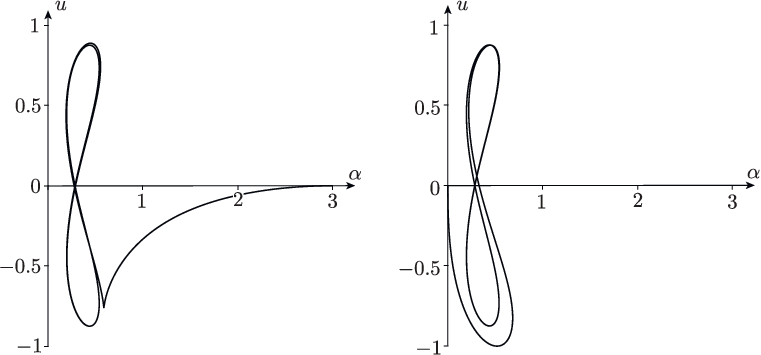}
	\end{center}
	\caption{A typical view of a periodic solution and
of trajectories tending to it for the system~(\ref{2812}) with $\lambda(\tau)=\sin\tau$, $\mu=0.3$.}
	\label{fig04}
\end{figure}

In the presence of friction force the reduced system can be reduced, after a transformation
similar to~(\ref{zz}), to the form
\begin{equation}
\begin{array}{l}
\label{eq5_K}
\alpha'=A u^2 - \sigma_1 \alpha, \quad u'=-\alpha u-\lambda(\tau)-\sigma_2 u,\\[3mm]
\displaystyle\frac{d \varphi}{d\tau}=u, \quad  \frac{d X}{d\tau}=\left( \frac{\alpha}{A} + a u \right)  \cos\varphi, \quad
\frac{d Y}{d\tau}= \left( \frac{\alpha}{A} + a u \right)   \sin\varphi,
\end{array}
\end{equation}
where $\sigma_1$ and $\sigma_2$~are the coefficients of friction.  As numerical experiments
show (see Fig.~\ref{fig04}), there is no acceleration in~(\ref{eq5_K}) in this case.

\begin{figure}[!ht]
	\begin{center}
		\includegraphics[totalheight=6cm]{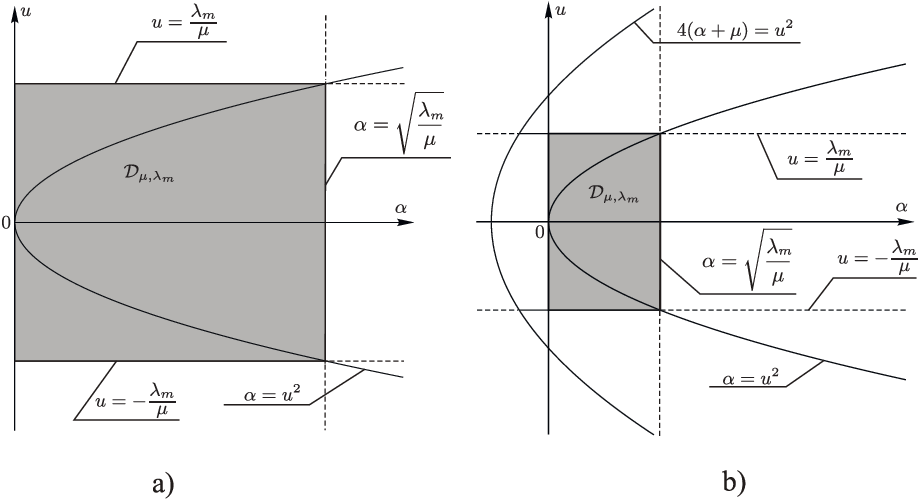}
	\end{center}
	\caption{}
	\label{fig08}
\end{figure}

Let us perform normalizing transformations of the variables, functions and time
$$
\alpha\to \sigma_1 \alpha, \quad u\to \frac{\sigma_1}{\sqrt{A}}u
$$
$$
\lambda\to \frac{\sigma_1^2}{\sqrt{A}}\lambda, \quad \sigma_2\to \sigma_1\mu, \quad \sigma_1 d\tau\to d\tau,
$$
 and represent the system as
\begin{equation}
\label{2812}
\alpha'=u^2-\alpha, \quad u'=-\alpha u-\lambda(\tau)-\mu u.
\end{equation}
We now apply the Brower theorem to this system (see Appendix А). To do so, we consider
a closed region $\mathcal{D}_{\mu, \lambda_m}$ on the plane
$(\alpha, u)$ bounded by four straight lines (see Fig.~\ref{fig08})
\begin{equation}
\begin{array}{ll}
\label{eq2912}
\displaystyle \gamma_1: \alpha=0, \quad \gamma_2: \alpha=\sqrt{\frac{\lambda_m}{\mu}}, \\   [4 mm]
\displaystyle \gamma_3: u=\frac{\lambda_m}{\mu}, \quad \gamma_4: u=-\frac{\lambda_m}{\mu},
\end{array}
\end{equation}
where $\lambda_m=\max |\lambda(\tau)|$.

We show that this region is invariant under the flow. Indeed, according to the second equation
in~(\ref{2812}), the vector field on the curves $\gamma_3$ and $\gamma_4$ is directed into the
region $\mathcal{D}_{\mu, \lambda_m}$ (since the inequalities $\dot{u}\leqslant
0$ and $\dot{u}\geqslant 0$ are satisfied for all $t$ on the upper straight line $\gamma_3$
and on the lower straight line $\gamma_4$, respectively). Similarly, on the straight line
$\gamma_1$ the vector field is also directed into the region (since on this straight line
$\dot{\alpha}\geqslant 0$), on the other hand, on the straight line $\gamma_2$ the vector field is also directed
inside (since the segment under consideration gets into the parabola $\alpha=u^2$,
where $\dot{\alpha}\leqslant 0$).

Hence, on the basis of Theorem~\ref{teoB1} of Appendix А we obtain the following theorem.

\begin{teo}
	In the system~$(\ref{2812})$ with $\mu\neq0$ there exists at least one periodic solution
in the region $\mathcal{D}_{\mu, \lambda_m}$.
\end{teo}

Now, in order to clarify the conditions for uniqueness of the periodic solution in the
region $\mathcal{D}_{\mu, \lambda_m}$, we find a region
on the plane $(\alpha, u)$ where the map for a period of the system~(\ref{2812}) is compressing.
According to Appendix А, this requires ascertaining the region of negative definiteness of
the symmetric part of the Jacobian of the right-hand side of the system~(\ref{2812}):
\begin{equation*}
\begin{array}{cc}
{\bf B}=
\left(\begin{array}{cc}
-1 & \frac{u}{2} \\
\frac{u}{2} & -\alpha-\mu
\end{array}\right).
\end{array}
\end{equation*}
Its eigenvalues $b_\pm$ are given by
$$
b_\pm=-\frac{1}{2}\big(\alpha+\mu+1\pm \sqrt{u^2+(\alpha+\mu-1)^2}\big).
$$
We thus find that the ``compression region'' (i.\,e., the region where $b_\pm<0$) is given by
$$
4(\alpha+\mu)\geqslant u^2.
$$
Using the equations of the upper and lower boundaries~(\ref{eq2912}) of the
region $\mathcal{D}_{\mu,\lambda_m}$, we obtain

\begin{teo}
	\label{Teo2}
	If in the system~(\ref{2812})
	\begin{equation}
	\label{eq}
	2\mu^{3/2}\geqslant \lambda_m,
	\end{equation}
	then in the region $\mathcal{D}_{\mu,\lambda_m}$ there exists the only periodic orbit to
which all other trajectories
	in $\mathcal{D}_{\mu,\lambda_m}$ tend exponentially.
\end{teo}

We note that in this section we obtain an estimate of the region of existence of the limit
cycle $\mathcal{D}_{\mu,\lambda_m}$
from the point of view of its localization. Generally speaking,
one can pose the problem of more exact localization of the limit cycle,
that is, the problem of finding the smallest possible region
$\mathcal{D}'\subset\mathcal{D}_{\mu,\lambda_m}$
containing this cycle.
On the other hand, under the conditions of Theorem \ref{Teo2} one can also pose the problem of
finding the largest possible region $\mathcal{D}''\supset\mathcal{D}_{\mu,\lambda_m}$
in which all trajectories tend to the only limit cycle.

The system~(\ref{eq5_K}) in the case $\sigma_2=0$ and $\lambda(\tau)=\sin\tau$ was considered
in~\cite{Tal}. In this case, inequality~(\ref{eq}) does not hold. However, numerical experiments
show that Theorem~\ref{Teo2} remains valid.

\begin{figure}[!ht]
	\begin{center}
		\includegraphics[totalheight=6.5cm]{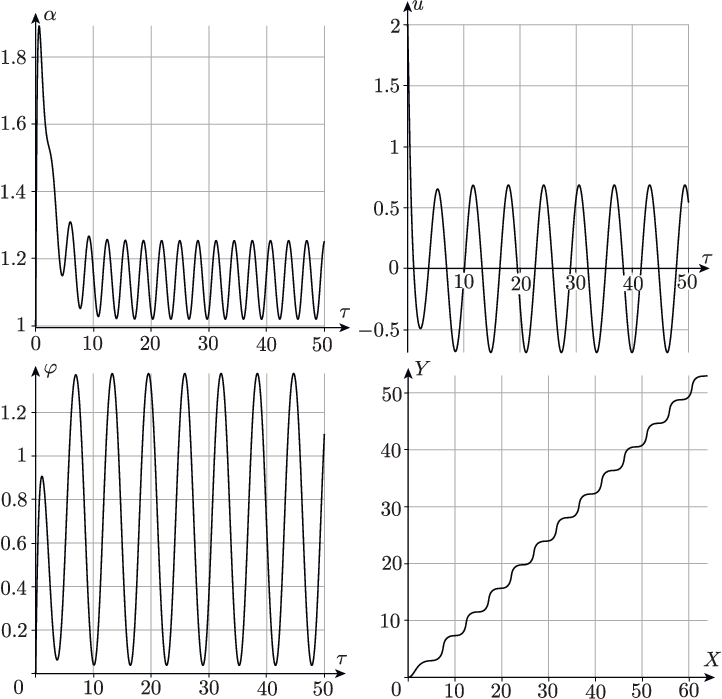}
		\caption{Functions $\alpha(\tau)$, $u(\tau)$, $\varphi(\tau)$, $Y(X)$ plotted
numerically for fixed parameters $\lambda_0=1$, $A=1$, $\sigma_1=0.1$, $\sigma_2=0$ and
initial conditions $\tau=0$, $\alpha=1$,  $u=2$,
			$\varphi=0$, $X=0$, $Y=0$.}
		\label{fig61}
	\end{center}
\end{figure}

{\bf Dynamics in configuration space.} A typical behavior of solutions to the
system~(\ref{eq5_K}) is shown in Fig.~\ref{fig61}.
We see that the ``limit'' motions of the point of contact, as in the absence of friction force,
are oscillations (with constant amplitude)
in a neighborhood of a straight line. In this case, in a neighborhood of the limit cycle the rotation
angle and the translational and angular velocities of the sleigh change periodically
with time (i.\,e., there is no unbounded increase in the translational velocity).

\begin{figure}[!ht]
	\begin{center}
		\includegraphics[totalheight=4.8cm]{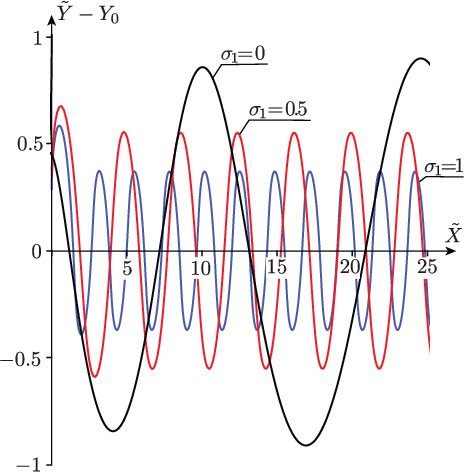}
		\caption{Trajectory of the point of contact plotted numerically for fixed parameters
$\lambda_0=1$, $A=1$, $\sigma_2=0$ and
			 initial conditions $\tau=0$, $\alpha=0$, $u=0$, $\varphi=0$, $X=0$, $Y=0$ for
different $\sigma_1$.}
		\label{fig65}
	\end{center}
\end{figure}

The trajectory of the point of contact in the variables\footnote{ Since the angle $\varphi(\tau)$
in a neighborhood of the limit cycle changes periodically with time, it suffices to set
$n=1$ in~(\ref{eq_phi}).} of
Section~\ref{DK} is presented in Fig.~\ref{fig65}.
This implies that, as the friction coefficient increases, the oscillation amplitude decreases.

\newpage

\section{The problem of acceleration in nonholonomic systems}

We discuss a number of problems that can be investigated by the methods presented in this paper.
The hydrodynamical model of the Chaplygin sleigh was proposed in~\cite{Fedorov}. Although this model
requires additional justification from the hydrodynamical point of view,
it would be interesting to explore a nonautonomous analogue of this model.
The same can be said of the problem of a sleigh with a constraint
inhomogeneous in the velocities, which has been investigated recently in~\cite{BM}.

Further we consider several systems of nonholonomic mechanics
with parametric excitation which is achieved by a given
periodic motion of some structural elements. Their common feature is that
the system of equations has a subsystem that governs the
evolution of (generalized) velocities with coefficients
periodically depending on time and can be considered as a separate set  (the dimension of this subsystem
does not exceed 2).

\begin{figure}[!ht]
	\begin{center}
		\includegraphics{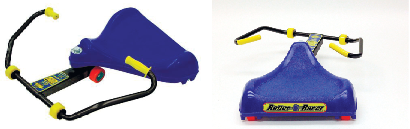}
		\caption{}
		\label{SR}
	\end{center}
\end{figure}

The simplest case is the {\it Roller Racer} \cite{Krishnaprasad}.
We recall that the {\it Roller Racer} consists of two coupled
bodies with a pair of wheels attached on each of the
bodies (see Fig. \ref{SR}).
The distinctive feature of the {\it Roller Racer} is that the
user moves
forward by oscillating the front handlebars in the transverse direction
from side to side.
As a rule, when investigating this system one assumes that, as a result of
external action, the angle between the two coupled bodies is
a given function of time (kinematic control).

In this case, one can decouple a linear equation with periodic coefficients
which governs the evolution of the translational velocity of one of
the bodies:
\begin{equation}
\label{eq1.1}
\dot{w}=A(t)w+B(t), \quad A(t)=A(t+T), \quad B(t)=B(t+T).
\end{equation}

For some particular mass distribution of the coupled bodies
equation~(\ref{eq1.1}) has been obtained in~\cite{Krishnaprasad},
where it is shown that in this case acceleration is observed. However,
this has not been rigorously proved as yet.

In a more complicated situation, namely, the Suslov problem with
moving masses, we have a special two-dimensional (nonlinear)
system that generalizes the system dealt with in this paper:
\begin{equation}
\begin{array}{ll}
\label{eq1.2}
\big({\bf G}(t)\boldsymbol w+\boldsymbol K(t)\big)^{\!\boldsymbol\cdot}=\Phi (\boldsymbol w){\bf J}\boldsymbol w, \\      [3 mm]
\boldsymbol w=(w_1, w_2), \quad
{\bf J}=
\left(
\begin{array}{cc}
0 & -1 \\
1 & 0
\end{array} \right),
\quad {\bf G}(t)={\bf G}^T(t) \\    [3 mm]
\Phi(\boldsymbol w)=a_1(t)w_1+a_2(t)w_2+k_3(t),
\end{array}
\end{equation}
where all functions of time $T$~are periodic and the symmetric matrix ${\bf G}(t)$ is positive definite for all $t$.

In the case where all coefficients of the system~(\ref{eq1.2}) are constant, its trajectories on the plane $(w_1, w_2)$
are very simple. First, there is an invariant submanifold~--- the straight line
\begin{equation}
\label{eq1.3}
a_1 w_1+a_2 w_2+k_3=0,
\end{equation}
which is filled with fixed points. The other trajectories lie on ellipses which are level lines of the energy integral
\begin{equation}
\label{eq1.4}
E=\frac{1}{2}\big(\boldsymbol w, {\bf G}\boldsymbol w\big)={\rm const}.
\end{equation}
The level lines~(\ref{eq1.4}) that intersect the straight
line~(\ref{eq1.3}) consist of two asymptotic trajectories connecting
a pair of equilibrium points, whereas the level lines that do not intersect
the straight line~(\ref{eq1.3}) define the periodic orbits of
the system~(\ref{eq1.2}).
Moreover, in this case, by a natural linear transformation
$$
u=a_1w_1+a_2w_2, \quad \alpha=A\big((a_1G_{12}-a_2G_{11})w_1+(a_1G_{22}-a_2G_{12})w_2\big), \quad A={\rm const}
$$
the system~(\ref{eq1.2}) is brought to the simple form
\begin{equation}
\label{eq1.5}
\dot{\alpha}=Au(u+k_3), \quad \dot{u}=-\frac{\alpha(u+k_3)}{A \det {\bf G}}.
\end{equation}

\newpage
\section*{Appendix A. The principle of compressing maps}
\addcontentsline{toc}{section}{Appendix A}
\setcounter{equation}{0}
\renewcommand{\theequation}{{\rm A}.\arabic{equation}}
\setcounter{teo}{0}
\renewcommand{\theteo}{{\bf A}.\arabic{teo}}

{\bf 1.} In this section, we briefly formulate in a form convenient for our purposes
some results on the periodic solutions of the system which
depend periodically on time. Let the following system be given in the Euclidean space
$E^n=\big\{\boldsymbol x=(x_1, \ldots, x_n)\big\}$:
\begin{equation}
\label{eqa1}
\boldsymbol{\dot{x}}=\boldsymbol v(\boldsymbol x, t), \quad \boldsymbol v(\boldsymbol x, t+T)=\boldsymbol v(\boldsymbol x, t).
\end{equation}

We first define the notion of an invariant subset $\mathcal{D}\subset \mathcal{M}$.
\begin{Def*}
If for all initial conditions $t_0\in [0, T]$ and $\boldsymbol x_0 \subset \mathcal{D}$
the trajectories $\boldsymbol x(t)\subset \mathcal{D}$, $t>t_0$, then $\mathcal{D}$~is said to be an
invariant
subset.
\end{Def*}

If the vector field~(\ref{eqa1}) has no singular points inside $\mathcal{D}$, then
a natural family of {\it maps for a period} is defined:
\begin{equation}
\label{eqA2}
\Pi_{t_0}: \mathcal{D}\to \mathcal{D},
\end{equation}
which assigns to each point $\boldsymbol x_0$ point $\boldsymbol x$ at time $t_0+T$, on the trajectory of the
system with initial conditions $\boldsymbol x(t_0)=\boldsymbol x_0$. The existence and uniqueness theorem and
the theorem of continuous dependence on initial conditions guarantee that the maps
$\Pi_{t_0}$ are continuous and biunique on their image.

If the set $\mathcal{D}$ is homeomorphic to a closed ball, then, according to the Brower
theorem, any map~(\ref{eqA2}) has a fixed point inside $\mathcal{D}$:
$$
\Pi_{t_0}(\boldsymbol x^*)=\boldsymbol x^*.
$$
In the flow~(\ref{eqa1}) this fixed point corresponds to the periodic solution
$$
\boldsymbol x_p(t)=\boldsymbol x_p(t+T), \quad \boldsymbol x_p(t_0)=\boldsymbol x^*.
$$
Thus, the following theorem holds.
\begin{teo}
	\label{teoB1}
	Suppose that the system~$(\ref{eqa1})$ admits an invariant set $\mathcal{D}$ that is homeomorphic
to a closed ball and does not contain any singular points of the vector field. Then in
$\mathcal{D}$ there is at least one periodic solution.
\end{teo}

{\bf 2.} Now assume that the flow~(\ref{eqa1}) admits a closed invariant set $\mathcal{D}$ inside
which it possesses the property of uniform compression, namely: the following inequality is satisfied
for any two trajectories
$\boldsymbol x_1(t), \, \boldsymbol x_2(t)$ inside $\mathcal{D}$ at all instants of time
$t$:
\begin{equation}
\label{eqA3}
|\boldsymbol x_1-\boldsymbol x_2|^{\boldsymbol \cdot}=\frac{\big(\boldsymbol x_1-\boldsymbol x_2, \, \boldsymbol v(\boldsymbol x_1, t)-\boldsymbol v(\boldsymbol x_2, t)\big)}{|\boldsymbol x_1-\boldsymbol x_2|}\leqslant -h|\boldsymbol{x}_1-\boldsymbol{x}_2|,
\end{equation}
where $h>0$~is some constant.

In this case the maps~(\ref{eqA2}) are also compressing
\begin{equation}
\label{eqA3_4}
|\Pi_{t_0}(\boldsymbol x_1)-\Pi_{t_0}(\boldsymbol x_2)|\leqslant \lambda|\boldsymbol x_1-\boldsymbol x_2|, \quad \lambda=e^{-hT}<1.
\end{equation}
Indeed, let $\Delta(t)=|\boldsymbol{x}^{(1)}(t) - \boldsymbol{x}^{(2)}(t) |$, where
$\boldsymbol{x}^{(k)}(t)$, $k=1,2$ are trajectories of the system~(\ref{eqa1}) that satisfy
the initial conditions
$\boldsymbol{x}^{(k)}(t_0)=x_k$, $k=1,2$. Then $\Delta(t_0)=|\boldsymbol{x}_1 - \boldsymbol{x}_2|$ and
$\Delta(t_0 + T)=|\Pi_{t_0}(\boldsymbol{x}_1)~-~\Pi_{t_0}(\boldsymbol{x}_2)|$.
From the above inequality~(\ref{eqA3}) and the condition $\Delta(t)>0$ we obtain
$$
\big(\ln \Delta(t) \big)^{\boldsymbol \cdot}\leqslant-h.
$$
Integrating this relation for a period, we obtain~(\ref{eqA3_4}).

Thus, we apply to $\Pi_{t_0}$ the principle of compressing Banach maps~\cite{Zehnder},
according to which $\Pi_{t_0}$ has a unique fixed point $\boldsymbol x^*$ in $\mathcal{D}$ to
which the trajectory of any other point tends exponentially.
Thus, the following theorem holds.
\begin{teo}
	\label{teoB2}
	Suppose that the system~$(\ref{eqa1})$ possesses the property of uniform compression
inside some closed invariant set
	$\mathcal{D}$. Then in $\mathcal{D}$ there exists a unique periodic solution and all
other trajectories in $\mathcal{D}$ tend to it exponentially.
\end{teo}

We now give the simplest criterion for uniform compression as presented in~\cite{Demidovich}.
Define the Jacobian of the right-hand side of~(\ref{eqa1}):
$$
{\bf J}(\boldsymbol x, t)=\bigg\|\frac{\partial v_i(\boldsymbol x, t)}{\partial x_j}\bigg\|.
$$

\begin{pro*}(\cite{Demidovich})
	Suppose that the quadratic form given by the matrix $\bf J$ is
uniformly negative definite for all $\boldsymbol x\in \mathcal{D}$ and
	$t$:
	$$
	\big(\boldsymbol x, {\bf J}(\boldsymbol x, t)\boldsymbol x\big)\leqslant -h(\boldsymbol x, \boldsymbol x),
	$$
	where $h>0$ is some constant. Then the flow of the system possesses the property
of uniform compression.
\end{pro*}
Finally, we obtain the following result.

\begin{teo}
If the eigenvalues of the matrix
$$
{\bf B}(\boldsymbol x, t)=\frac{1}{2} \big({\bf J}(\boldsymbol x, t)+{\bf J}^T(\boldsymbol x, t)\big), \quad \boldsymbol x\in \mathcal{D}
$$
are negative and separated from zero, then the system~$(\ref{eqa1})$
possesses a unique limit cycle in $\mathcal{D}$ to which all other trajectories tend exponentially.
\end{teo}

\section*{Acknowledgments}

The authors extend their gratitude to S.\,P.~Kuznetsov and D.\,V.~Treschev for valuable discussions and comments.

This work was supported by the Russian Science Foundation (project 14-50-00005).

\end{document}